\documentclass[twocolumn,prb,aps,floatfix]{revtex4}

\usepackage{graphicx}
\usepackage{dcolumn}
\usepackage{bm}

\begin{document}

\preprint{}

\title{Dependence of the electronic structure of self-assembled
  InGaAs/GaAs quantum dots on height and composition}

\author{Gustavo A. Narvaez}
\affiliation{National Renewable Energy Laboratory, Golden, Colorado 80401}
\author{Gabriel Bester}
\affiliation{National Renewable Energy Laboratory, Golden, Colorado 80401}
\author{Alex Zunger}
\affiliation{National Renewable Energy Laboratory, Golden, Colorado 80401}

\date{\today}

\begin{abstract}
  While electronic and spectroscopic properties of self-assembled In$_{\rm
    1-x}$Ga$_{\rm x}$As/GaAs dots depend on their shape, height and alloy
  compositions, these characteristics are often not known accurately from
  experiment. This creates a difficulty in comparing measured electronic and
  spectroscopic properties with calculated ones. Since simplified theoretical
  models (effective mass, k$\cdot$p, parabolic models) do not fully convey the
  effects of shape, size and composition on the electronic and spectroscopic
  properties, we offer to bridge the gap by providing accurately calculated
  results as a function of the dot height and composition. Prominent features
  of our results are the following. (i) Regardless of height and composition,
  the confined electron energy levels form shells of nearly degenerate states
  with a predominant orbital ``s'', ``p'', $\cdots$ character. In contrast,
  the confined hole energy levels form shells only in flat dots and near the
  highest hole level (HOMO). (ii) In alloy dots, the {\em electrons}' ``s-p''
  splitting depends weakly on height, while the ``p-p'' splitting depends
  non-monotonically---due to alloy fluctuations. In pure, non-alloyed
  InAs/GaAs dots, {\em both} these splittings depend weakly on height.
  Further, the ``s-p'' splitting is larger while the ``p-p'' has nearly the
  same magnitude. For {\em holes} levels in alloy dots, the ``s-p'' splitting
  decreases with increasing height (the splitting in tall dots being about 4
  times smaller than in flat dots), whereas the ``p-p'' splitting remains
  nearly unchaged. Shallow, pure non-alloyed dots have a ``s-p'' splitting of
  nearly the same magnitude, whereas the ``p-p'' splitting is about three
  times larger. (iii) As height increases, the ``s'' and ``p'' character of
  the wavefunction of the HOMO becomes mixed, and so does the heavy- and
  light-hole character. (iv) In alloy dots, regardless of height, the
  wavefunction of low-lying hole states are localized inside the dot.
  Remarkably, in non-alloyed InAs/GaAs dots these states become localized at
  the interface as height increases. The localized states are nearly
  degenerate and polarized along $[1\bar 10]$ and $[110]$.  This localization
  is driven by the peculiarities of the biaxial strain present in the
  nanostructure.
\end{abstract}

\maketitle

\section{Introduction}

The electronic structure and spectroscopic properties of quantum dots,
including
excitons,\cite{exciton_SK,charged.X_SK,finley_PRB_2001,multiexcitons_SK,dekel_PRL_1998,finestructure.X_SK}
charged excitons,\cite{charged.X_SK,finley_PRB_2001}
multiexcitons,\cite{finley_PRB_2001,multiexcitons_SK,dekel_PRL_1998} and
excitonic fine-structure,\cite{finestructure.X_SK} all depend on the size and
shape of the dots. This dependence reflects both quantum confinement effects,
as well as shape-induced band-folding and inter-band
coupling.\cite{zunger_pss.a_2002} {\em Simulations} of electronic structure
and spectroscopic properties of quantum
dots\cite{he_PRB_2004,shumway_PRB_2001,williamson_PRB_2000,bester_PRB_2003a,bester_PRB_2003b,simonin_PRB_2004,wojs_PRB_1996}
must naturally assume the size and shape of the dot. On the other hand, {\em measurements} of spectroscopic properties of a
dot\cite{exciton_SK,charged.X_SK,finley_PRB_2001,multiexcitons_SK,dekel_PRL_1998,finestructure.X_SK}
are rarely accompanied by accurate measurements of the size and shape of the GaAs-covered dot,
except in rare cases where detailed cross-sectional scanning tunneling
microscopy experiments are performed, such as in Refs.
\onlinecite{liu_APL_2002,bruls_APL_2002,gong_APL_2004}.
This situation creates a significant difficulty, if not a crisis, in
interpreting spectroscopic data on quantum dots, and in critically testing
various theoretical approaches. Thus, in reality one is often forced to
address the inverse problem,\cite{shumway_PRB_2001} namely, fit the
spectroscopic data to a theoretical model by using the size and shape as
adjustable parameters. The difficulty with this approach is threefold: First,
in this approach all theories, no matter how na\"ive, ultimately work by
virtue of forcing a fit to experiment, even if the assumptions entering the
theory may seem unjustified in their own right (e.g. assuming single-band
effective mass; neglect of strain; neglect of spin-orbit coupling.) Second,
there are usually too many free-parameters, which involve not only non-trivial
shapes, but also unknown composition profiles (e. g., In$_{{\rm 1-x}}$Ga$_{\rm
  x}$As dots).  Third, since the relationship between shape and spectroscopic
properties is model dependent, unrealistic shapes are often deduced. For
example, in simple effective-mass,\cite{wojs_PRB_1996,jacak_book,ema_refs}
k$\cdot$p,\cite{kdotp_refs} or parabolic models the ``p'' and ``d'' levels of
electrons and holes are degenerate if one assumes spherical, lens-shaped,
cubic, or cylindrical dots. In contrast, in more advanced {\em atomistic}
models---like empirical
pseudopotential\cite{wang_PRB_1999,wang_APL_2000,zunger_pss.b_2001} or tight
binding\cite{TB_method}---those levels are split even for the above mentioned
ideal shapes, resulting in clear spectroscopic signatures. To fit measured
spectroscopic signatures of actual dots by simple theoretical models, one
needs to assume at the outset irregular shapes. For instance, Dekel {\em et
  al.}\cite{dekel_PRL_1998} needed to assume a parallepipedal box to explain
their multiexciton data on non-alloyed InAs/GaAs dots; and Ferreira assumed
shape distortions to explain fine structure.\cite{ferreira_PhysicaE_2002} Such
assumptions are not needed in atomistic approaches to modeling.

In this work we have used a high-level atomistic approach to predict the
spectroscopic characteristics of In$_{\rm 1-x}$Ga$_{\rm x}$As/GaAs
self-assembled dots as a function of the most crucial geometric parameter,
namely the heigh. We calculate strain profiles, ``p-'' and ``d-''level
splittings as well as electron and hole wavefunctions. While ultimately it
will be necessary for experimentalists to report the size, shape, and
composition profile to which their spectroscopic data correspond, the type of
study reported here may be used to bridge, in the interim, spectroscopy with
theory without clouding the issue by severe theoretical approximations.
Similar studies were carried out by the group of Bimberg in Refs.
\onlinecite{grundmann_PRB_1995,stier_PRB_1999,guffarth_PRB_2001,shape_x.xx_kp},
by Shumway {\em et al.} in Ref. \onlinecite{shumway_PRB_2001}, by Williamson
{\em et al.} in Ref.  \onlinecite{williamson_PRB_2000}, by Kim {\em et al.}
in Ref.  \onlinecite{kim_PRB_1998} and by Pryor in Ref.
\onlinecite{pryor_PRB_1999}.

%
\section{Methods and their illustration}

\subsection{Choice of dot geometries}

Among several geometries, Stranski-Krastanow quantum dots growth in lens
shape.\cite{lens-shaped_dots} Hence, we focus on lens-shaped, self-assembled
In$_{\rm 1-x}$Ga$_{\rm x}$As/GaAs quantum dots (QDs). In addition, we include
a 1-monolayer-thick In$_{\rm 1-x}$Ga$_{\rm x}$As wetting layer (WL). Figure 1
shows a sketch of the geometry of the nanostructure. (QD+WL+GaAs matrix). We
focus on dots with $x=0.4$ and pure InAs. The QDs have circular base with
diameter $b=252\,${\AA} and height $h$ in the interval
$20\,${\AA}-$75\,${\AA}.

%
\begin{figure}
\includegraphics[width=7.5cm]{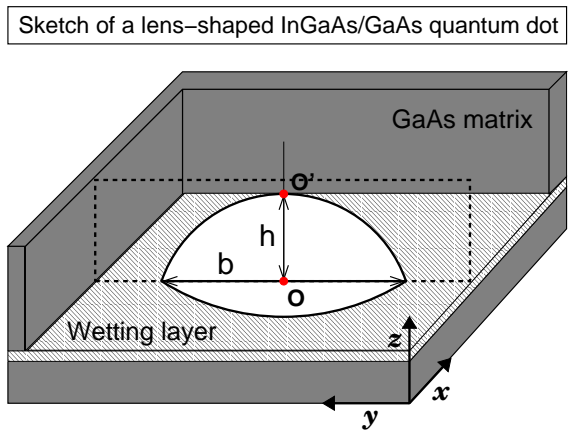}
\caption{{\label{Fig_1}}Sketch of the geometry of self-assembled In$_{\rm
    1-x}$Ga$_{\rm x}$As/GaAs quantum dots. A section of the GaAs matrix (grey)
  has been ommited for clarity. The dot (white) is lens shaped with base
  diameter $b$ and height $h$.  The dot lies on top of a 1-monolayer-thick
  In$_{\rm 1-x}$Ga$_{\rm x}$As wetting layer. Two auxiliary features are
  shown: (i) A plane that bisects the dot (dashed lines) and (ii) the line
  $\stackrel{\longleftrightarrow}{OO^{\prime}}$, which pierces the dot through
  the center along the [001] direction.}
\end{figure}

\subsection{Strain relaxation done via atomistic, not continuum elasticity}
\label{strain.app}

The position-dependent strain profiles present in the nanostructure are
usually addressed within (harmonic) {\em continuum} elasticity (CE)
theory.\cite{tadic_JAP_2002,pryor_JAP_1998} Notwithstanding, strain profiles
can also be calculated within atomistic elasticity (AE) in the form of a
valence force field.\cite{original_VFF} In AE the strain energy is expressed
in terms of atomic positions and interatomic force constants. While the force
constants are fit to the elastic constants $C_{11}$, $C_{12}$ and $C_{44}$,
much like in CE, expressing the strain in terms of atomic positions contains
much more information than expressing strain in terms of CE coordinates. For
example, a square-based pyramid has a $C_{4v}$ point group symmetry if
described by CE. This means that the $[110]$ and $[\bar 110]$ facets are taken
to be equivalent. In this case, all ``p'' energy levels are degenerate.
However, in AE the point group symmetry of a square-based pyramid made of InAs
is $C_{2v}$. In this case, the [110] and $[\bar 110]$ facets are
symmetry-inequivalent and ``p'' levels are split.

In general, three interrelated sources are responsible for the splitting of
the ``p'' states.\cite{bester_PRB_2005} A quantum dot with a base that
globally possesses inversion symmetry in the (001) plane---like a lens, a
truncated pyramid or a truncated cone---has in reality the lower $C_{2v}$
symmetry (for pure, non-alloyed InAs/GaAs dots), which originates from the
underlying zinc-blende atomic lattice. Thus, the {\em first source} is the
lack of inversion symmetry in the $C_{2v}$ point group, which manifests itself
around the dot-matrix interfaces where the [110] and [1$\bar{1}$0] directions
are inequivalent.  The {\em second source} is the propagation of the strain
field towards the center of the dot as a consequence of the atomic relaxation.
The {\em third source} is the piezoelectric effect (which magnitude is
presently under debate\cite{bester_PRB_2005}) that arises from the strain
field of $C_{2v}$ symmetry. It should be noted that approximations used in
previous calculations of the piezoelectric effect
\cite{prev_piezo,pryor_PRB_1997} have been shown to be crude and further
investigations have been called for.\cite{bester_PRB_2005}

Our choice of AE is based on a generalization of Keating model to 3
terms---bond-bending, bond stretching and their cross
terms.\cite{williamson_PRB_2000} We fit the elastic constants $C_{11}$,
$C_{12}$ and $C_{44}$ of zinc-blende InAs and GaAs, as well as the correct
dependence of the Young's modulus with pressure for both materials. In the
In$_{\rm 1-x}$Ga$_{\rm x}$As alloy system, the bond-stretching and the
bond-stretching/bond-bending cross-term parameters for the mixed cation
Ga-As-In bond angle are taken as the algebraic average of the In-As-In and
Ga-As-Ga values. While the ideal bond angle is $109^{\rm o}$ for the pure
zinc-blende crystal, to satisfy Vegard's law for the alloy volume, the value
of $110.5^{\rm o}$ was used for the mixed bond angle. AE is superior to CE in
that it does not assume harmonicity---in fact, anharmonic effects can be
explicitly included\cite{lazarenkova_APL_2004} into the valence force
field.\cite{original_VFF}

Here, we calculate the position-dependent strain tensor
$\widetilde{\varepsilon}({\bf R})$ within the atomistic elasticity approach.
To calculate the strain tensor cubic components $\varepsilon_{ij}$ ($i,j=x$,
$y$, $z$), we proceed in two steps: (1) We relax the atomic positions within
the supercell in order to minimize the elastic energy, which is given by a
generalized valence force field.\cite{williamson_PRB_2000} (2) We relate the
relaxed (equilibrium) atomic positions with the unrelaxed atomic positions via
strain tensor.\cite{pryor_JAP_1998} At each equilibrium position ${\bf R}_{l}$
of atom $l$ we identify the tetrahedron formed by its 4 nearest neighbors.
This tetrahedron is distorted in comparison to the unrelaxed tetrahedron.
Thus, the three edges of these tetrahedra that are determined by the vectors
connecting the four neighbors can be related by the strain tensor
$\widetilde{\varepsilon}$ as follows.

\begin{widetext}
\begin{equation}
\label{strain.eq}
\left(
\begin{array}{ccc}
 {\bf R}_{al,x} & {\bf R}_{bl,x} & {\bf R}_{cl,x} \\
 {\bf R}_{al,y} & {\bf R}_{bl,y} & {\bf R}_{cl,y} \\
 {\bf R}_{al,z} & {\bf R}_{bl,z} & {\bf R}_{cl,z}
\end{array}
\right)= 
\left(
\begin{array}{ccc}
 1+\varepsilon_{xx} & \varepsilon_{yx} & \varepsilon_{zx} \\
 \varepsilon_{xy} & 1+\varepsilon_{yy} & \varepsilon_{yz} \\
 \varepsilon_{xz} & \varepsilon_{yz} & 1+\varepsilon_{zz}
\end{array}
\right)
\times
\left(
\begin{array}{ccc}
 {\bf R}^0_{al,x} & {\bf R}^0_{bl,x} & {\bf R}^0_{cl,x} \\
 {\bf R}^0_{al,y} & {\bf R}^0_{bl,y} & {\bf R}^0_{cl,y} \\
 {\bf R}^0_{al,z} & {\bf R}^0_{bl,z} & {\bf R}^0_{cl,z}
\end{array}
\right)
,
\end{equation}
\end{widetext}

\noindent where ${\bf R}_{al}$, ${\bf R}_{bl}$, and ${\bf R}_{cl}$ are the 3
vectors that connect, respectively, neighbors 1 and 2, 2 and 3, and 3 and 4 in
the equilibrium, distorted tetrahedron that encloses atom $l$. ${\bf
  R}^0_{al}$, ${\bf R}^0_{bl}$, and ${\bf R}^0_{cl}$ are the corresponding
vectors (edges) in the unrelaxed tetrahedron. From Eq. (\ref{strain.eq}), the
cubic strain tensor components are calculated by a matrix inversion.

%
\begin{figure*}
\includegraphics[width=17.0cm]{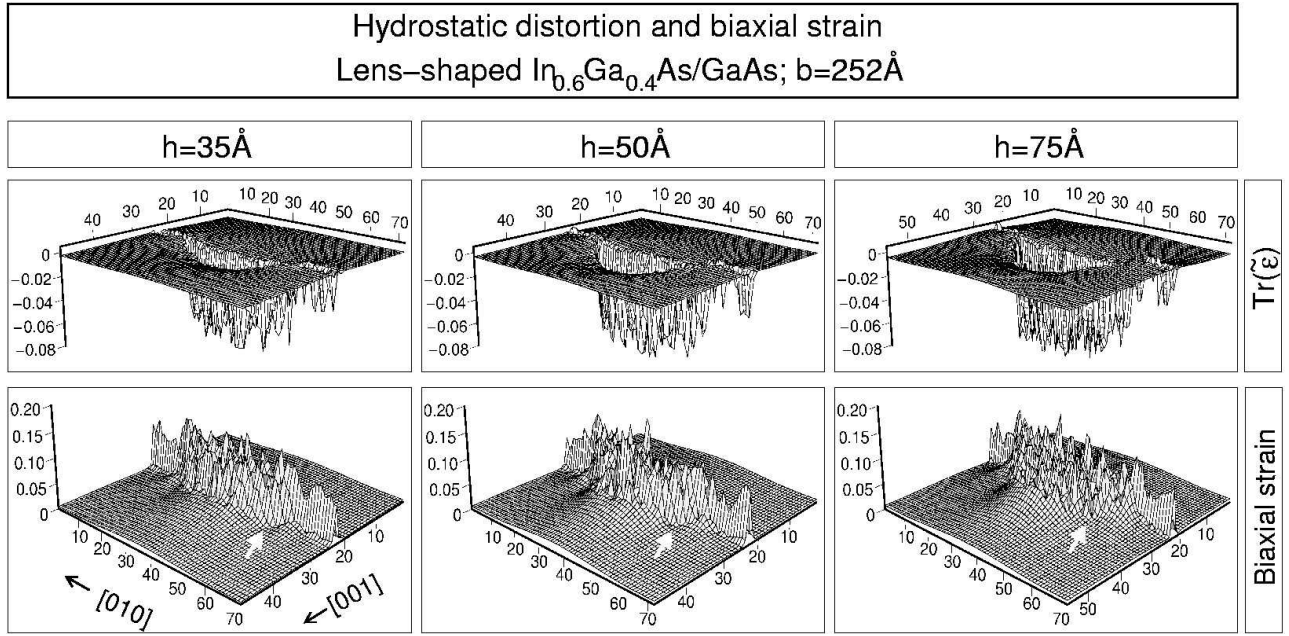}
\caption{{\label{Fig_2}} Hydrostatic distortion [${\rm
  Tr}(\tilde{\varepsilon})$; top] and biaxial strain (bottom) as a
  function of dot's height plotted on a planar section
  that is normal to [100] and bisects the dot (see Fig. \ref{Fig_1}).
  Positions are measured in units of $a_{GaAs}=5.65\,${\AA}. The wetting layer
  (WL) strain appears prominently. Alloy fluctuations make the strain profiles
  sample dependent.  Two robust features emerge: (i) The biaxial strain at the
  top of the dot increases with height and slowly decays off. The taller the
  dot the faster this decay. ${\rm Tr}(\tilde{\varepsilon})$ shows a similar
  trend. (ii) The underlying zincblende atomic structure causes the strain to
  rapidly drop to zero along the ${\rm [011]}$ and ${\rm [01\bar{1}]}$
  directions. Arrows indicate this feature.}
\end{figure*}

Figure \ref{Fig_2} shows the biaxial strain

\begin{equation}
B=\sqrt{(\varepsilon_{xx}-\varepsilon_{yy})^2+(\varepsilon_{yy}-\varepsilon_{zz})^2+(\varepsilon_{zz}-\varepsilon_{xx})^2}
\end{equation}

\noindent and the hydrostatic distortion (isotropic strain) $I={\rm
  Tr}(\tilde{\varepsilon})$ as a function of dot height. The results presented
correspond to the strain values on a planar section in the [100] direction
that bisects the dot (see Fig. \ref{Fig_1}). The spikes in the dot and wetting
layer region are a consequence of the random arrangement of In and Ga atoms.
Thus, the spikes structure significantly changes from dot to dot. Two
additional features are also prominent. (i) The biaxial strain at the
dot-matrix interface, located near the top of the dot, increases with height
and slowly decays off. This decay is faster as the dot becomes taller. The
latter can be understood by considering the dot as a spherical cap inscribed
in a sphere of radius $\rho_0=[h^2+(b/2)^2]/2h$ (the {\em taller} the dot, the
{\em smaller} $\rho_0$) and by assuming that the decay of $B$ outside the dot
is qualitatively well described by $(\rho_0/\rho)^{3}$, which is the radial
dependence of the biaxial strain outside a continuum, elastic sphere of radius
$\rho_0$ embedded in a lattice-mismatched medium.\cite{elshelby_JAP_1954}
Here, $\rho$ is the radial distance from the sphere surface.  In our
simulations, ${\rm Tr}(\tilde{\varepsilon})$ shows a similar trend as $B$.
(ii) Due to the underlying zincblende atomic structure, the biaxial strain
drops to zero along the $[011]$ and $[01\bar 1]$ directions, as well as along
their equivalent crystallographic directions. (See white arrows in Fig.
\ref{Fig_2}.)  This feature is robust upon changes in height, being present
for all the quantum dots we considered.

\subsection{The single-particle electronic structure is calculated via pseudopotential plane-wave method instead of ${\bf k}\cdot {\bf p}$}

To calculate the energies and wavefunctions of electron and hole states in the
quatum dot, we use the empirical pseudotential method of Wang and
Zunger.\cite{wang_PRB_1999} This approach combines a pseudopotential
description of the single-particle Hamiltonian with the linear combination of
bulk bands (LCBB) method to solve for the energies and
wavefunctions.\cite{wang_PRB_1999} In this method, the Hamiltonian reads

\begin{equation}
\label{hamiltonian.lcbb}
{\cal{H}}=-\frac{\beta}{2}\nabla^2+V^{SO}+\sum_{\alpha={\rm
    In,Ga,As}}\sum_{l}\,v_{\alpha}({\bf R}-{\bf R}^{\alpha}_{l};\widetilde\varepsilon),
\end{equation}

\noindent where $\beta$ is an empirical parameter that accounts for
non-locality effects; $V^{SO}$ is a non-local empirical operator that
describes the spin-orbit interaction;\cite{williamson_PRB_2000}
$v_{\alpha}({\bf \eta};\widetilde\varepsilon)$ is a screened pseudopotential
(for atom of type $\alpha$) that depends on strain; and ${\bf R}^{\alpha}_{l}$
is the relaxed vector position of atom $l$ of type $\alpha$. The dependence of
the atomic pseudopotential on strain transfers to the electronic Hamiltonian
the information on atomic displacements. The strain-dependent pseudopotential
reads 

\begin{equation}
v_{\alpha}({\bf R}-{\bf R}^{\alpha}_{l};0)[1+\gamma_{\alpha}\,{\rm
  Tr}(\widetilde{\varepsilon})], 
\end{equation}

\noindent where $\gamma_{\alpha}$ is a fitting
parameter. It should be noted that $v_{\alpha}({\bf R}-{\bf
  R}^{\alpha}_{l};\widetilde\varepsilon)$ is fit to {\em bulk} properties of
GaAs and InAs, including bulk band structures, experimental deformation
potentials and effective masses, and LDA-determined band offsets. In order to
improve the transferability of pseudopotential $v_{\alpha}({\bf R}-{\bf
  R}^{\alpha}_{l};\widetilde\varepsilon)$, a simple dependence on the chemical
environment of atom $\alpha$ is introduced. For instance, for $\alpha=As$ in
an environment of $p$ Ga atoms and $p-4$ In atoms we use

\begin{equation}
v^{(p)}_{As}=\frac{(4-p)}{p}\,v_{As}({\rm InAs})+\frac{p}{4}\,v_{As}({\rm GaAs}).
\end{equation} 

\noindent The pseudopotentials used in this work have been successfully tested for quantum
wells.\cite{williamson_PRB_2000}

The wavefunction of state $i$ is $\psi_{i}({\bf R})$, which satisfies ${\cal
  H}\psi_i={\cal E}_i\psi_i$, is expanded in bulk Bloch states $u^{(M)}_{n,{\bf
    k}}({\bf R})$ of material $M$. [It should be noted that the bulk materials
($M$) can be strained.\cite{wang_PRB_1999}] Namely,

\begin{equation}
\label{psi.lcbb}
\psi_i({\bf R})=\sum_{M}\sum_{n,{\bf k}}\,C^{\,(i)}_{M;n,{\bf
    k}}\,\left[\frac{1}{\sqrt{N}}\,u^{(M)}_{n,{\bf k}}({\bf R})e^{i{\bf k}\cdot{\bf R}}\right],
\end{equation}

\noindent where, $n$ and ${\bf k}$ indicate the band-index and wave-vector of
the Bloch state, respectively; and $N$ is the number of primary cells
contained in a supercell that encloses the quantum dot. Thus, by diagonalizing
the Hamiltonian [Eq. (\ref{hamiltonian.lcbb})] in the Bloch states basis we
find the coefficients $C^{\,(i)}_{M;n,{\bf k}}$. The calculated wavefunctions
$\psi_i({\bf R})$ are 2-fold, Kramers degenerate, so we have omitted the spin
index $\sigma$. This representation for the wave function is very different
from the familiar k$\cdot$p method in that in the latter approach the basis
set is constructed only from states near the Brillouin zone center ($\Gamma$),
while here we use a full-zone description. Further, in k$\cdot$p one is
restricted to just one or two Bloch bands at $\Gamma$ ($8\times 8$
representating two Bloch bands) while here we consider $n$ Bloch bands.  In
Appendix \ref{levels.app}, we present an assessment of the convergence of
energy levels ${\cal E}_i$ as a function of the expansion parameters in Eq.
(\ref{psi.lcbb}).

\subsection{Strain-modified band offsets}
\label{strain.offsets}

\begin{figure*}
\includegraphics[width=13.25cm]{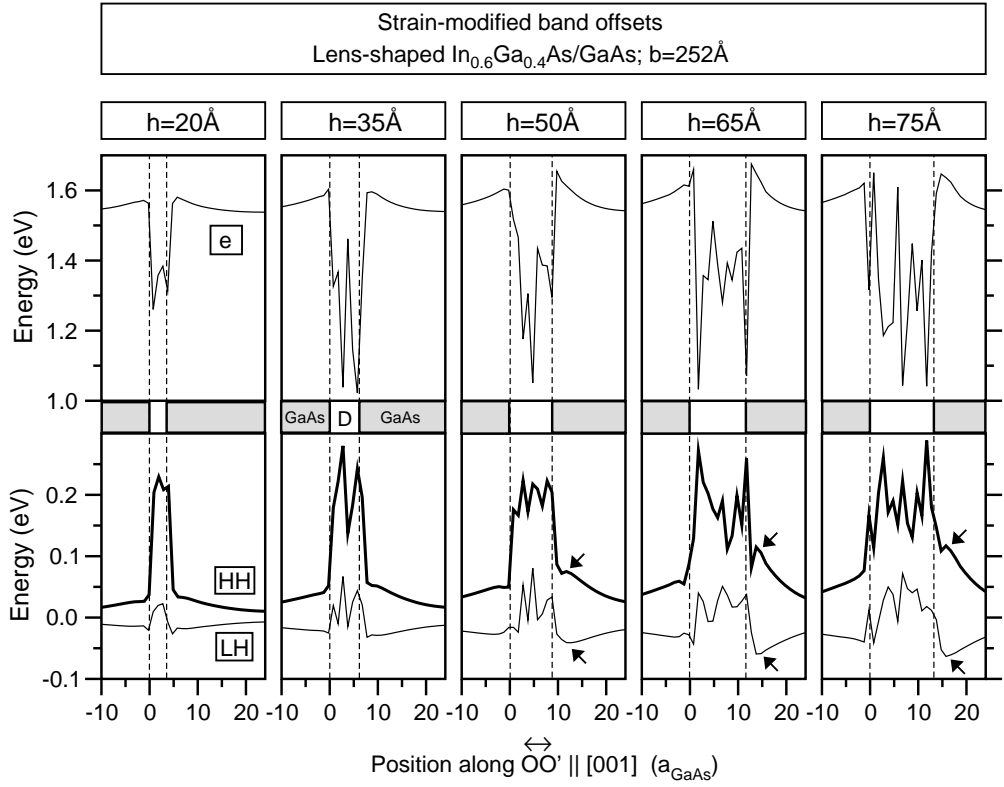}
\caption{{\label{Fig_3}} Strain-modified band offsets for electrons (upper
  panel, $e$), and first and second holes (lower panel; thick and thin lines,
  respectively) along the
  $\stackrel{\longleftrightarrow}{OO^{\prime}}\,\parallel\,[001]$ as a
  function of height. (See Fig. \ref{Fig_1}.) In the dot region (indicated
  with a $D$) the offsets are irregular (jagged) due to alloy fluctuations.
  These irregularities are more noticeable as the height increases. Outside
  $D$ the heavy-hole (HH) and light-hole (LH) character reverses. Arrows
  indicate ``pocket'' formation in the second hole band offset as the dot's
  height increases.}
\end{figure*}

While we describe the effects of strain on the electronic structure
atomisticallyas shown in Eq. \ref{hamiltonian.lcbb}, here, for {\em
  illustrative purposes only}, we calculate the conduction (electron) and
valence (hole) strain-modified band offsets (confining potentials) present in
the quantum dot by coupling strain to k$\cdot$p-like equations. At each 8-atom
unit cell in the supercell we diagonalize the conduction and valence
(including spin-orbit coupling) band strain Hamiltonians that Wei and Zunger
put forward in Ref. \onlinecite{wei_PRB_1994}.  Figure \ref{Fig_3} shows the
calculated strain-modified band offsets for electrons and holes along the line
$\stackrel{\longleftrightarrow}{OO^{\prime}}$ indicated in Fig. \ref{Fig_1}.
Electron band offset appears in the upper panel, whereas thick and thin lines
in the lower panel show the first and second holes offsets, respectively. We
note the following features. (i) The band offsets (Fig. \ref{Fig_3}) inherit
the jagged nature of the strain fields (Fig. \ref{Fig_2}). Inside the quantum
dot (region ``D'' in Fig.  \ref{Fig_3}), alloy fluctuations lead to a small
mixing in the heavy-hole (HH) and light-hole (LH) character of the band
offsets.  (ii) Regardless of the dot height, the higher energy hole band
offset has HH character inside the dot and LH outside. Conversely, the lower
energy hole band offset has LH character inside and HH outside.  (iii) The
increase of biaxial strain at the dot-matrix interface that occurs as the
height increases (Fig. \ref{Fig_2}) is reflected in the increase (decrease) of
the higher (lower) energy hole band offset. (See arrows in Fig.  \ref{Fig_3}).
In particular, for tall dots, the decrease of the lower energy hole offset at
the interface leads to the formation of a {\em pocket} in the band offset.

%

\subsection{Orbital and Bloch character of wavefunctions}

The single-band effective mass model predicts that in the continuum limit, in
which lens-shaped dots have full axial symmetry, the confined energy levels
will form shells of degenerate states with {\em definite} angular momentum
$l=0,\,1,\,2,\,\cdots$.  \cite{harmonic_osc}Thus, the {\em orbital} character
of the confined levels will be {\em pure} ``s'', ``p'', ``d'', etc. 
However, when the correct symmetry of the dot is taken in to account, as it is
the case when using our atomistic pseudopotential method, the confined energy
levels can have mixed orbital character. To quantify the degree of mixing, we
analyse the single-particle wavefunctions $\psi_i({\bf R})$ by projecting
their {\em envelopes} on eigenfunctions of the axial ($|| z$, see Fig.
\ref{Fig_1}) angular momentum $e^{i\,m\phi}/\sqrt{2\pi}$ and determine the
orbital (angular) character. The latter given by

\begin{equation}
\label{angular.char}
A^{(i)}_{m,n}=\int dz\int d{\bf \rho}\rho \left|\int d\phi\,f^{(i)}_n({\bf \rho},\phi,z)\frac{e^{i\,m\phi}}{\sqrt{2\pi}}\right|^2.
\end{equation}

\noindent In Eq. (\ref{angular.char}), we have used cylindrical coordinates,
i. e. ${\bf R}=(\rho,\phi,z)$, and written the $\Gamma$-derived envelope
    function as 

\begin{equation}
f^{(i)}_n(\rho,\phi,z)=\sum_{M}\sum_{{\bf
    k}}\,\sum_{n^{\prime}}C^{\,(i)}_{M;n,{\bf k}}\langle u^{(M)}_{n,{\bf
    k}}|u^{(M)}_{n^{\prime},\Gamma}\rangle\,\frac{e^{i{\bf k}\cdot{\bf R}}}{\sqrt{N}}.
\end{equation} 

The Bloch character of the hole wavefunctions $\psi^h_i({\bf R})$ in the
quantum dot depends on strain as well as band coupling. To study the extent to
which the Bloch character is HH, LH, and SO, we proceed similarly and project
the envelope of $\psi^h_i({\bf R})$ on the total-angular-momentum basis
$|J,J_z\rangle$ ($\{|3/2,\pm 3/2\rangle,|3/2,\pm 1/2\rangle,|1/2,\pm
1/2\rangle\}$).

\subsection{Exciton energy levels are obtained via the screened configuration interaction approach, not perturbation theory}
\label{CI.model}

We calculate the exciton energy levels $E_{\nu}$ by using the configuration
interaction method as proposed in Ref.  \onlinecite{franceschetti_PRB_1999}.
Briefly, in this method, the exciton states

\begin{equation}
\Psi_{\nu}=\sum_{i,j}C^{\,(\nu)}_{ij}|e_{i}h_{j}\rangle,
\end{equation} 

\noindent where $\{|e_{i}h_{j}\rangle\}$ denotes a basis of single-substitution Slater
determinants (configurations), in which an electron is promoted from
$\psi^{(h)}_j({\bf R})$ to $\psi^{(e)}_i({\bf R})$. The CI method would
deliver the exact exciton ground and excited states in the case of a complete
basis.  However, we {\em truncate} this basis and consider all the possible
configurations build out of $n_e$ electron and $n_h$ hole confined states. The
coefficients $C^{\,(\nu)}_{ij}$ arise from the diagonalization of the exciton
Hamiltonian.  The direct ($J_{ij}={\Gamma}^{\,i,j}_{j,i}$) and exchange
($K_{ij}=\Gamma^{\,i,j}_{i,j}$) electron-hole Coulomb integrals that enter the
calculation are derived from the Coulomb scattering matrix elements

\begin{widetext}
\begin{equation}
{\Gamma}^{\,i,j}_{k,l}=e^2\int\int {\rm d}{\bf R}{\rm d}{\bf R}^{\prime}
\frac{\Big[\psi^{(h)}_i({\bf R})\Big]^{*}\Big[\psi^{(e)}_j({\bf R}^{\prime})\Big]^{*}\psi^{(e)}_{k}({\bf R}^{\prime})\psi^{(h)}_{l}({\bf
    R})}{\epsilon({\bf R},{\bf
    R}^{\prime})|{\bf R}-{\bf R}^{\prime}|} 
\end{equation}
\end{widetext}

\noindent The microscopic, phenomenological dielectric constant $\epsilon({\bf R},{\bf
  R}^{\prime})$ that screens the interaction is calculated within the
Thomas-Fermi model proposed by Resta.\cite{resta_PRB_1977} We do not use
simple perturbation theory where the exciton is described via a
Coulomb-corrected single-particle band gap $({\cal E}^e_i-{\cal
  E}^h_j)-J_{ij}$; here, ${\cal E}^e_i$ and ${\cal E}^h_j$ are the energies of
electron level $i$ and hole level $j$, respectively.

%
%
\section{Results}

We now present the effects of height and composition on the energies of
confined levels and their splittings, wavefunctions of selected confined
electron and hole levels, and the lowest transition energy of the exciton.

\subsection{Energies of confined levels: ``p'' levels split even for ideal lens-shape dots}

\begin{figure}
\includegraphics[width=8.5cm]{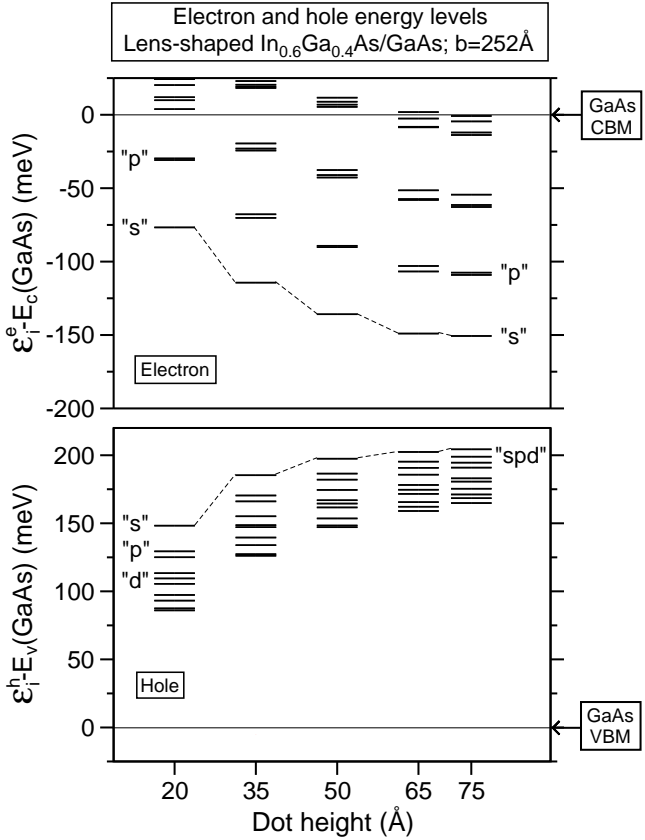}
\caption{{\label{Fig_4}}First 20 electron (top panel) and
  hole (bottom) energy levels ${\cal E}^e_n$ and ${\cal E}^h_n$ for In$_{\rm
    0.6}$Ga$_{\rm 0.4}$As/GaAs dots with different heights. The electron and
  hole energies are measured from GaAs CBM [$E_c(GaAs)=-4.093\,{\rm eV}$] and
  GaAs VBM [$E_v(GaAs)=-5.620\,{\rm eV}$], respectively. As height increases,
  the number of confined levels (${\cal E}^e_n<0$ and ${\cal E}^h_n>0$)
  increases and the single-particle gap (${\cal E}^e_0-{\cal E}^h_0$)
  decreases. The confined electronic energy levels group in {\em
    non-degenerate shells} for all dots. ``s'', ``p'', and ``d'' indicate the
  predominant orbital character of selected states. For holes, up to
  $h=50\,${\AA} the second and third levels are ``p''-like, while for larger
  heights these levels are ``s''-``p'' hybridized.}
\end{figure}

\begin{figure}
\includegraphics[width=8.5cm]{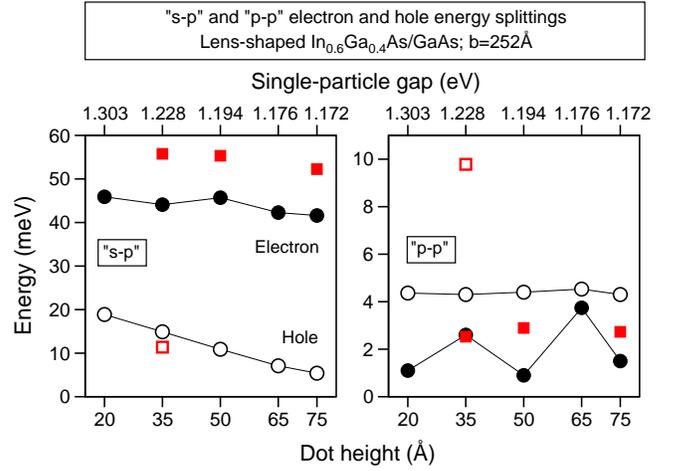}
\caption{{\label{Fig_5}}Height dependence of the ``s-p'' (${\cal E}_1-{\cal
    E}_0$) and ``p-p'' (${\cal E}_2-{\cal E}_1$) energy splittings for
  electron and hole levels in In$_{\rm 0.6}$Ga$_{\rm 0.4}$As/GaAs. The
  single-particle gap is also shown. The ``s-p'' splitting for electrons
  is bigger than for holes; in turn, the ``p-p'' splitting shows the opposite
  trend. The height dependence of the ``p-p'' splitting for electrons is not
  monotonic, due to alloy fluctuations, while for holes this splitting remains
  nearly unchanged. For comparison, we show the ``s-p'' and ``p-p'' splittings
  for electron (solid squares) and hole (open squares) in non-alloyed
  InAs/GaAs dots.}
\end{figure}

The confined electron and hole energy levels are respectively labeled as
${\cal E}^{e}_i$ and ${\cal E}^h_i$, where $i=0,\; 1, \cdots$ is an orbital
quantum number. The corresponding wavefunctions are $\psi^{(e)}_n({\bf r})$
and $\psi^{(h)}_n({\bf r})$.  Each of the confined states are 2-fold, Kramers
degenerate. We label the $i=0$ electron and hole states as LUMO and HOMO,
respectively. The first 20 electron and hole energy levels of an In$_{\rm
  0.6}$Ga$_{\rm 0.4}$As/GaAs dot appear in Figure \ref{Fig_4}. The electron
and hole energies are measured, respectively, from the bottom of the
conduction band (CBM) $E_c(GaAs)=-4.093\,{\rm eV}$ (calculated bulk electron
affinity) and from the top of the valence band (VBM) $E_v(GaAs)=-5.620\,{\rm
  eV}$ (calculated bulk ionization potential) of bulk GaAs.  For all heights,
the confined electron states form groups ({\em shells}) of {\em
  non-degenerate} levels.\cite{harmonic_osc} In turn, this shell structure can
only be identified for the first few hole levels (near HOMO) in shallow dots
(up to $h=50\,${\AA}). For taller dots, the holes show no shell structure.
The number of confined electron states is significantly smaller than that of
holes.  While for the tallest dot there are 10 confined electron states, more
than 150 hole states are confined in all the considered dots.  When comparing
the results for $h=35\,${\AA} with the first 20 electron and hole energy
levels (not shown) in a lens-shaped, non-alloyed InAs/GaAs dot, we find that
(i) the electron levels form shells that have a bigger average separation
($55\,{\rm meV}$ vs $45\,{\rm meV}$ in In$_{\rm 0.6}$Ga$_{\rm 0.4}$As/GaAs),
(ii) the hole energy level structure near the HOMO is significantly different
(see splittings in Fig. \ref{Fig_5}) and (iii) the single-particle gap ${\cal
  E}^e_0-{\cal E}^h_0$ is smaller (see Fig. \ref{Fig_9}).

%

{\em Energy splittings.---}The ``s-p'' energy splitting ${\cal E}_1-{\cal
  E}_0$ and ``p-p'' splitting ${\cal E}_2-{\cal E}_1$ for electrons and holes
are shown in Figure \ref{Fig_7} as a function of the dot height. (The dots'
single-particle gap is also indicated.) Two features emerge. First, ``s-p''
splitting for electrons is bigger than for holes. For electrons, the magnitude
of the splitting is about $45\,{\rm meV}$ and depends weakly on height. On the
other hand, for holes, this splitting changes from $\simeq 20\,{\rm meV}$ to
nearly $5\,{\rm meV}$ when height changes from $20\,${\AA} to $75\,${\AA}.
Second, the ``p-p'' splitting shows the opposite behavior. Namely, for holes
the magnitude of the splitting remains nearly constant at approximately
$4\,{\rm meV}$, and for electrons it changes non-monotonically. It should be
noted that the electronic ``p-p'' splitting is sensitive to alloy fluctuations
and it can change by almost a factor of two by changing the alloy realization
in the dot.\cite{narvaez_tobepublished}

For comparison, ``s-p'' and ``p-p'' splittings in a lens-shaped, non-alloyed
InAs/GaAs dots with $b=252\,${\AA} also appear in Fig. \ref{Fig_5}. The pure
dot has a ``s-p'' splitting for holes that is nearly the same as in the
$h=35\,${\AA} In$_{\rm 0.6}$Ga$_{\rm 0.4}$As/GaAs dot, while this splitting
for electrons is nearly 20\% bigger and depends weakly on height.  In
contrast, the ``p-p'' splitting for holes in the pure (non-alloyed InAs/GaAs)
dot is about twice as big as in the $h=35\,${\AA} alloy dot, and for electrons
the ``p-p'' splitting depends weakly on height and is similar in magnitude to
the splittings in alloy dots. It should be noted that in pure (non-alloyed
InAs/GaAs) dots the {\em hole} energy levels undergo a localization crossover
(see discussion in Sec.  \ref{local.strain}) for tall dots, which render
meaningless the notion of ``s-p'' and ``p-p'' splittings. (Such a localization
crossover is absent for electrons.)  For this reason, we have compared the
splittings for the holes in a shallow ($h=35\,${\AA}) non-alloyed InAs/GaAs
dot only, while comparing the splitting for electrons in a range of heights.

%

\subsection{Wavefunctions of confined states in In$_{0.6}$Ga$_{0.4}$As/GaAs
  and non-alloyed InAs/GaAs dots: mixed ``s''+``p'' character and ``HH''+``LH'' character}

Figure \ref{Fig_5} compares the wavefunctions of LUMO (lowest electron) and
HOMO (highest hole) states as a function of height in In$_{\rm 0.6}$Ga$_{\rm
  0.4}$As/GaAs dots with base $b=252\,${\AA}. To make the comparison, we plot
isosurfaces that enclose $75\%$ of the total charge density, and show contour
plots taken at $1\,nm$ above the dot's base. In addition, the ``{\em
  s}''-orbital character of the LUMO and HOMO is indicated as well as the
HOMO's heavy- and light-hole character. Further, for each height, the energy
${\cal E}^e_0$ of LUMO is shown relative to $E_c(GaAs)$ and the energy of HOMO
${\cal E}^h_0$ relative to $E_v(GaAs)$. Prominent results are the following.
{\em LUMO:} The lateral spatial extent of the wavefunction depends weakly on
height. In contrast, as height decreases, the wavefunction extends into the
barrier along the vertical direction. The ``{\em s}''-orbital character of
LUMO remains at nearly $90\%$.  {\em HOMO:} Wavefunctions are more sensitive
to height, showing a spatial extension that gets reduced significantly both in
the lateral and vertical direction when height changes from $75\,${\AA} to
$20\,${\AA}. This reduction leads to a strong localization at the center of
the dot for $h=20\,${\AA}. The ``s''-orbital character of the HOMO remains at
about 90\% up to $h=50\,${\AA}, for taller dots the character of the HOMO
becomes ``s'' and ``p'' mixed. This mixing reflects the reduction of the hole
charge density near the center of the dot. The heavy- (HH) and light-hole (LH)
character also change with height in a similar manner as the ``s'' and ``p''
character. Namely, for the three smaller dots ($20\,${\AA}, $35\,${\AA},
$50\,${\AA}) the HOMO is mostly of the HH type, but as the height increases
the LH character increases.
We expect this behavior since the LH band-offset increases within the dot as
height increases. (See Fig. \ref{Fig_3}.)

%
\begin{figure}
\includegraphics[width=8.5cm]{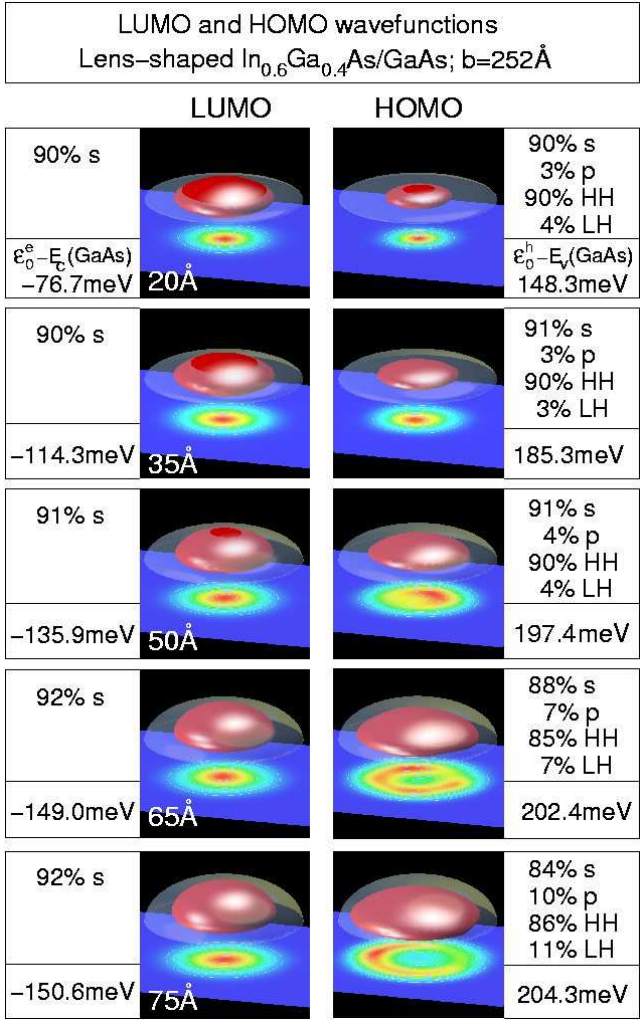}
\caption{{\label{Fig_6}} Height dependence of lowest electron level (LUMO) and
  highest hole level (HOMO) wavefunctions. Isosurfaces enclose 75\% of the
  total charge density. Countours are taken at a plane $1\,${\rm nm} above the
  base. The energy relative to $E_c(GaAs)$ and $E_v(GaAs)$ is shown. The
  ``s-'' and ``p-''orbital character (see Eq. \ref{angular.char}), and the heavy-hole (HH)
  and light-hole (HH) character are indicated. For the tallest dot, the
  wavefunctions are entirely confined within the dot and the effective size of
  the dot---spatial extent of the wavefunctions---is nearly the same for LUMO
  and HOMO. For shorter dots the effective size gets significantly reduced for
  the HOMO while it remains nearly unchanged for the LUMO.}
\end{figure}

\begin{figure}
\includegraphics[width=8.5cm]{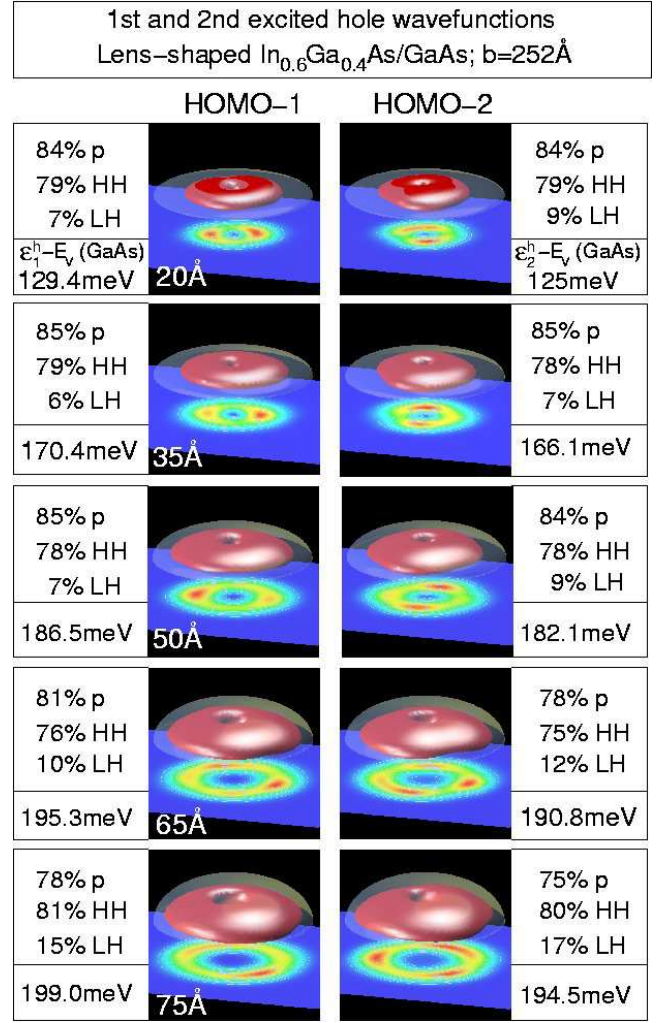}
\caption{{\label{Fig_7}}Wavefunctions of the first 2 excited hole levels
  HOMO-1 and HOMO-2 for different dot's height. As in Fig. \ref{Fig_6}, the
  isosurface encloses $75\%$ of the total charge density and the countours are
  taken at 1nm above the base. Labels indicate the ``p'' orbital character, the
  heavy- and light-hole characters and the energy of the states relative to
  $E_v(GaAs)$.}
\end{figure}

{\em Excited hole states.---}Figure \ref{Fig_7} compares wavefunctions for
excited hole states; namely, second (HOMO$-1$) and third (HOMO$-2$) hole
states. As in Fig.  \ref{Fig_5}, we show isosurfaces and contour plots,
present the orbital and HH/LH character of these states, and the energies
${\cal E}^h_1$ and ${\cal E}^h_2$ for HOMO$-$1 and HOMO$-$2, respectively. The
character of HOMO$-1$ and HOMO$-2$ are nearly the same at all heights. In
addition, these states have a dominant ``p''-orbital character for {\em all}
heights, regardless the absence of ``p''-shell structure for tall dots
($h=65\,${\AA} and $75\,${\AA}). (See Fig. \ref{Fig_4}) As in the case of the
HOMO state, the HOMO$-1$ and HOMO$-2$ states have increasing light-hole
character with increasing height.  However, the percentage of light-hole
character is almost twice that of HOMO.  For instance, at $h=75\,${\AA}, the
LH character of the HOMO$-2$ is 17\% while the HOMO LH character is 11\%.

%

\subsection{Strain-driven hole localization}
\label{local.strain}

\begin{figure}
\includegraphics[width=8.5cm]{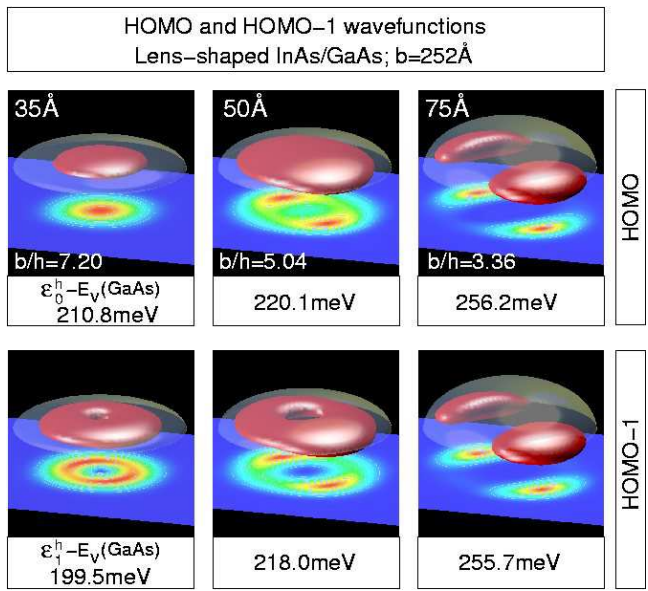}
\caption{{\label{Fig_8}}HOMO and HOMO-1 wavefunctions in pure, non-alloyed InAs/GaAs
  lens-shaped quantum dot as a function of height. The dots have the same base
  diamater $b=252${\AA}. The aspect ratio $b/h$ is shown, as well as the
  energy relative to $E_v(GaAs)$. As before, the isosurfaces enclose 75\% of
  the charge density and the countours are taken at $1\,{\rm nm}$ above the
  base. As height increases, interface localization takes plave, and the HOMO and
  HOMO$-$1 become nearly degenerate.}
\end{figure}

For dots containing both Ga and In (In$_{0.6}$Ga$_{0.4}$As/GaAs) with
$b=252\,${\AA} and heights in the range 20-75{\AA}, we have shown that the
wavefunctions of HOMO (${\cal E}^h_0$) and HOMO$-$1 (${\cal E}^h_1$), as well
as other low-lying excited hole states, are localized inside the dot (see
Figs.  \ref{Fig_6} and \ref{Fig_7}) and that these levels have a sizeable
``s-p'' splitting ${\cal E}^h_1-{\cal E}^h_0$ (see Fig. \ref{Fig_5}). In
contrast, in lens-shaped pure, non-alloyed InAs/GaAs quantum dots, localization of the
low-lying hole states at the dot interface develops as the height of the dot
increases. In addition, HOMO and HOMO$-$1 become nearly degenerate [e. g., ${\cal
  E}^h_0-E_v(GaAs)=256.2\,{\rm meV}$ and ${\cal E}^h_1-E_v(GaAs)=255.7\,{\rm
  meV}$ at $h=75${\AA}], as well as HOMO$-$2 and HOMO$-$3. (It should also be
noted that for the flat dot the energies of HOMO and HOMO$-$1 are bigger than
in the alloy dot of the same size.) Remarkably, HOMO
and HOMO-1 are {\em polarized} along $[1\bar 10]$ (Fig. \ref{Fig_8}) while
HOMO$-$2 and HOMO$-$3 (not shown) are polarized along $[110]$. Figure
\ref{Fig_8} shows the development of this interfacial localization and the
energy of HOMO and HOMO$-$1 relative to $E_v(GaAs)$.  As in Figs. \ref{Fig_6}
and \ref{Fig_7}, isosurfaces enclose 75\% of the HOMO and HOMO$-$1 total
charge density.

%

\begin{figure*}
\includegraphics[width=16.0cm]{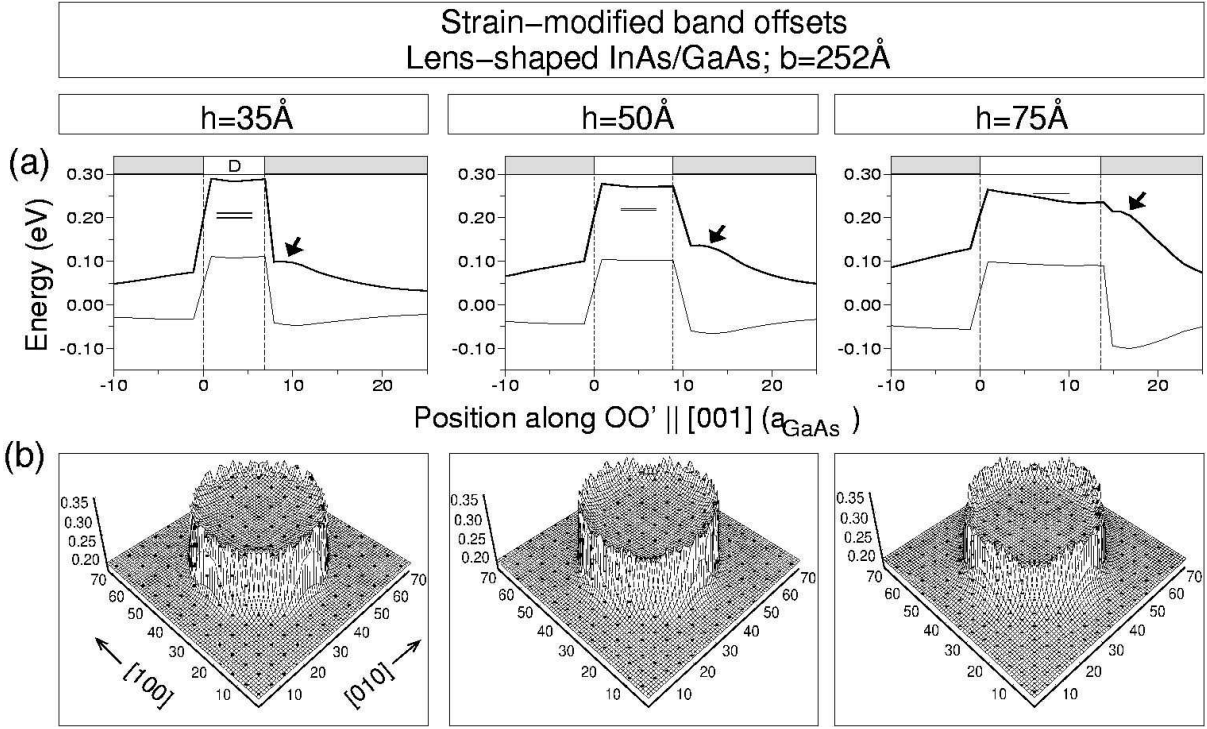}
\caption{{\label{Fig_8prime}}Strain-modified band offsets in a pure,
  lens-shaped, non-alloyed InAs/GaAs quantum dot along the
  $\stackrel{\longleftrightarrow}{OO^{\prime}}$ direction (a) and in a plane
  normal to latter cut at a height slightly above the base of the dot (b).
  Dashes in panel (a) correspond to the HOMO and HOMO$-$1 energy levels. The
  splitting among these levels decreases with the dot height. Arrows indicate
  pocket formation (see also Fig. \ref{Fig_3}) in the band offset due to
  strain accumulation at the dot-matrix interface at the top of the dot.}
\end{figure*}

It should be noted that the pure, non-alloyed InAs/GaAs dots have the {\em
  same} base size and {\em same} height range as the alloy dots we have
previously discussed.  Hence, hole localization does not have its origin on
geometrical aspects of the quantum dots. Instead, the localization of the hole
wavefunctions is driven by the biaxial strain present in the nanostructure
(QD+GaAs matrix). To explain this result, we plot in Fig. \ref{Fig_8prime}(a)
the valence band offsets of heavy-hole (HH) and light-hole (LH) character
(inside the dot) along the line $\stackrel{\longleftrightarrow}{OO^{\prime}}$
(see Figs.  \ref{Fig_1} and \ref{Fig_3}) and the energy of HOMO and HOMO$-$1
(thick dashes).  In addition, we plot in Fig. \ref{Fig_8prime}(b) a
3-dimensional rendering of the higher energy hole band offset values at a
plane normal to $\stackrel{\longleftrightarrow}{OO^{\prime}}$, cut slightly
above the base of the dot. {\em First}, we see that as the dot heigh increases
the pocket structure [indicated with arrows in Fig. \ref{Fig_8prime}(a)] that
appears in the higher energy band offset dramatically widens the confining
potential (given by the band offset) along
$\stackrel{\leftrightarrow}{OO^{\prime}}$ and that the band offsets values
decrease. Further, the band offset become assymetric with the values at the
top of the dot {\em smaller} that at the {\em base}. As a consequence, it
becomes clear why the energy of the HOMO and HOMO-1 become bigger and we also
expect the wavefunctions of these states to be localized near the base of the
dot. It is important to mention that in the calculation of the energy of the
HOMO and HOMO$-$1 we do not utilize the band offset we present in Fig.
\ref{Fig_8prime}(a). {\em Second}, the band offsets shown in Fig.
\ref{Fig_8prime} present a ``crown'' structure at the dot-matrix
interface.\cite{note-000} This crown structure becomes more significant as
height increases, as a consequence of the values of band offset along the
$[110]$ and $[1\bar 10]$ direction presenting a weak dependence on height.
{\em Finally}, the localization of the low-lying hole wave functions that
develops as the height of pure, non-alloyed InAs/GaAs lens-shaped quantum dots
is a results of the peculiar characteristics of the higher energy valence-band
offsets, which in turn are determined by the biaxial strain profile in the
nanostructure.

Finally, it should be noticed that electron states do not experience interface
localization; instead, they continue to form shells and the levels in these
shells have a predominant, not-mixed orbital character (``s'', ``p'',
$\cdots$). The reason for the absence of electron localization is that
electron states are sensitive to the (hydrostatic) isotropic strain rather
than to the biaxial strain.

%

\subsection{Exciton transitions}

\begin{figure}
  \includegraphics[width=7.5cm]{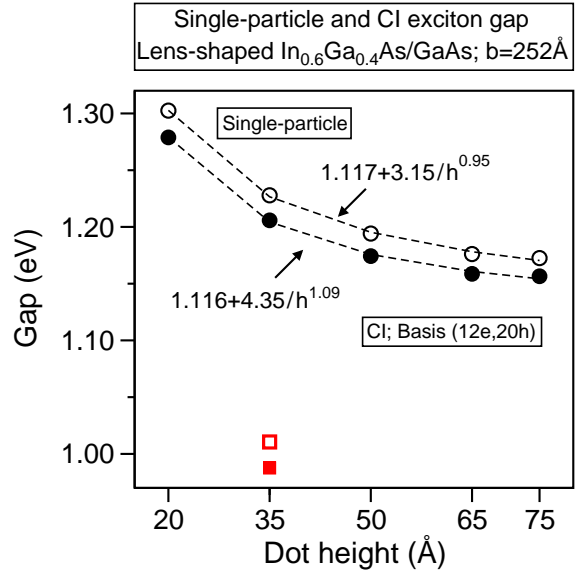}
\caption{{\label{Fig_9}}Exciton gap as a function of height. Single-particle
  (open circles) and configuration interaction (CI, solid circled) results are
  shown. The CI basis is $(n_e=12,n_h=20)$ for $h=35$-$75\,${\AA} and
  $(n_e=6,n_h=20)$ for $h=20\,${\AA}. Dashed lines represent fits to the
  function $a+b/h^{\gamma}$. Fitting parameters are indicated. For comparison,
  we present results [squares; open (SP) and solid (CI)] for a pure,
  non-alloyed InAs/GaAs dot.}
\end{figure}

Figure \ref{Fig_9} shows the lowest bright transition (gap) of the exciton as
a function of dot height, calculated at the single-particle (SP) level (${\cal
  E}^e_0-{\cal E}^h_0$) and by using the many-body, configuration interaction
(CI) method.\cite{franceschetti_PRB_1999} In the latter, we use a basis
consisting of 12 electron and 20 hole confined levels $(n_e=12,n_h=20)$ for
heights $h=35$-$75\,${\AA} and $(n_e=6,n_h=20)$ for $h=20\,${\AA}. As
expected, the transition energy decreases as height increases, due to
confinement. However, the scaling with height differs significantly from
predictions of single-band, effective mass ($\sim
h^{-2}$).\cite{gap_effective.mass} Namely, by fitting our results for the
height dependence of the gap to the function $a+b/h^{\gamma}$ we find
$\gamma^{SP}=0.95$ and $\gamma^{CI}=1.09$. The value of the gap for large
heights correspond, respectively, to $a^{SP}=1.117\,{\rm eV}$ and
$a^{CI}=1.116\,{\rm eV}$ in the single-particle and CI approach.  We expect
the SP and CI scaling exponents to be different, as in the single-particle
calculation the scaling is dictated by the scaling of the LUMO and HOMO,
whereas in the CI calculation the electron-hole matrix elements are also
included and the magnitude of these matrix elements decreases with height.
The discrepancy between the scaling ($\gamma$) in single-band effective mass
and both our SP and CI calculations can be attributed to non-trivial effects
that are naturally accounted for within our atomistic approach such as
non-parabolicity and multiband effects, and the position-dependent strain
present in the nanostructure. The values of $a$ at large heights is also
expected to be different, and the difference is attributed to correlation
effects that are present in the many-body, configuration approach. Thus, we
expect a small difference ($\sim {\rm meV}$) between $a^{SP}$ and $a^{CI}$.

For comparison, we also present results for the lowest exciton transition
[squares, open (SP) and solid (CI)] in a lens-shaped, non-alloyed InAs/GaAs
quantum dot. The values of the transition energies are smaller than in the
alloy dots.

%

\section{Summary}

By using a high-level atomistic approach, we have predicted spectroscopic
characteristics of self-assembled In$_{\rm 1-x}$Ga$_{\rm x}$As/GaAs quantum
dots as a function of height and composition. Several prominent features
emerged.

(i) The biaxial strain at the dot-GaAs matrix interface increases with height,
whereas, regardless of height, the strain is negligibly small along the
$[011]$, $[01\bar 1]$ and crystallographically equivalent directions.

(ii) Regardless of height and composition, the confined electron energy levels
group in shells of nearly degenerate states. The average energy splitting
among these shells depends weakly on height; however, this splitting is larger
in pure, non-alloyed InAs/GaAs. In contrast, the confined hole energy levels
form shells only in flat dots and near the highest hole level (HOMO).

(iii) In alloy dots, the {\em electrons}' ``s-p'' splitting depends weakly on
  height, while the ``p-p'' splitting depends non-monotonically---due to alloy
  fluctuations. In pure, non-alloyed InAs/GaAs dots, {\em both} these
  splittings depend weakly on height. Further, the ``s-p'' splitting is
  larger while the ``p-p'' has nearly the same magnitude. For {\em holes}
  levels in alloy dots, the ``s-p'' splitting decreases with increasing height
  (the splitting in tall dots being about 4 times smaller than in flat dots),
  whereas the ``p-p'' splitting remains nearly unchaged. Shallow pure dots
  have a ``s-p'' splitting of nearly the same magnitude, whereas the ``p-p''
  splitting is about three times larger.

(iv) As height increases, the ``s'' and ``p'' character of the wavefunction of
the HOMO becomes mixed, and so does the heavy- and light-hole character. 

(v) In alloy dots, regardless of height, the wavefunction of low-lying (near
the HOMO) hole states are localized inside the dot. Remarkably, in pure,
non-alloyed InAs/GaAs dots, as the dot height increases, these states become
localized at the dot-matrix interface and nearly degenerate. Further, the
localized states are polarized along $[1\bar 10]$ (HOMO and HOMO-1) and
$[110]$ (HOMO-2 and HOMO-3). This localization effect is driven by the
peculiarities of the biaxial strain present in the nanostructure.
 
(vi) The lowest exciton transition energy (gap) decreases with height, but the
scaling (roughly $\sim h^{-1}$) differs significantly from the prediction of
single-band effective mass ($\sim h^{-2}$).

The study we presented here may be used to bridge spectroscopy results with
theory without the need for severe theoretical approximations.

\begin{acknowledgments}
This work has been supported by DOE-SC-BES-DMS under grant No. DE-AC36-99GO10337.
\end{acknowledgments}

\appendix
\section{Convergence} 
\label{levels.app}

The LCBB method contains several convergence parameters: supercell size,
number of bulk materials ($M$), number of bands ($n$), and number of ${\bf k}$
points. The choice of bulk materials and number of bands can be physically
motivated and justified, whereas the choice of supercell size and number of
{\bf k} points is not as clear. For this reason, we conducted a converge
assesment on supercell size and number of {\bf k} points. As for the bulk
materials, we choose (i) {\em unstrained} GaAs and (ii) InAs subject to strain
values of $\varepsilon_{xx}=\varepsilon_{yy}=-0.06$, and
$\varepsilon_{zz}=+0.04$. These values are typical of strain inside a
non-alloyed InAs/GaAs lens-shaped quantum dot. For the bands $n$, when
simulating the {\em electron} and {\em hole} energy levels we use respectively
the lowest {\em conduction} band and the three highest valence bands of both
bulk materials.

%
\begin{figure*}
\includegraphics[width=14.5cm]{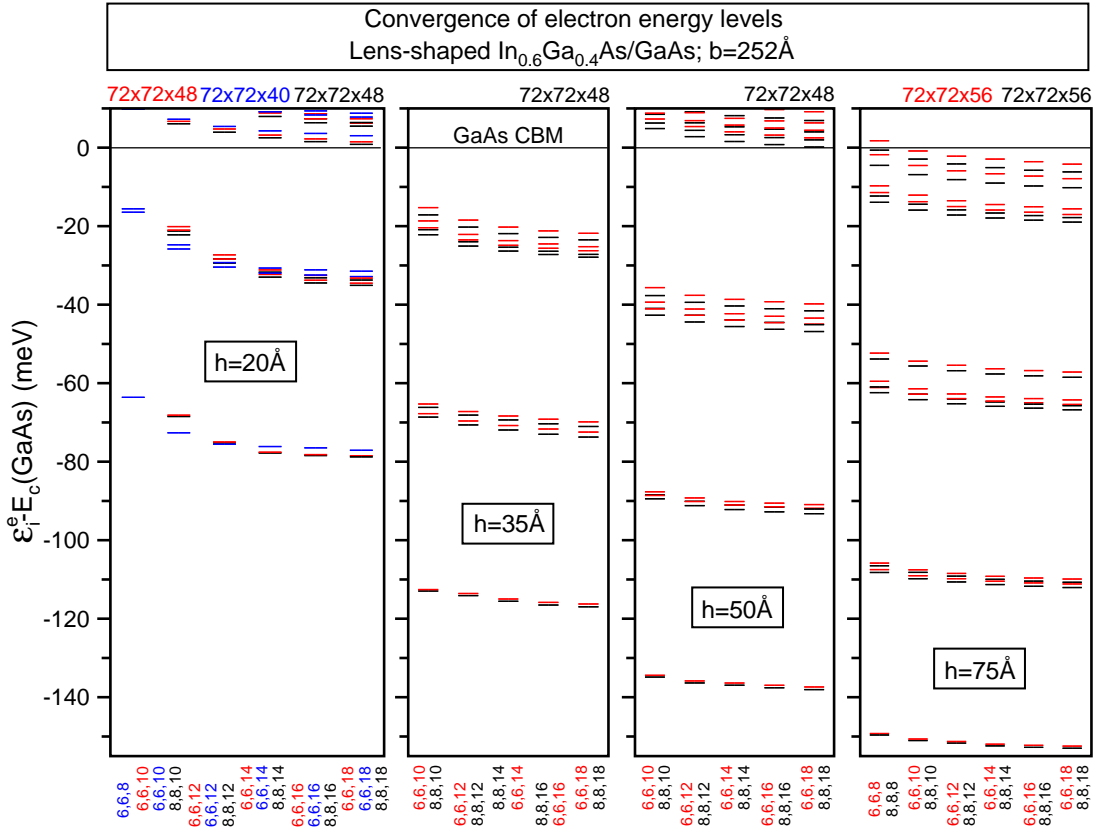}
\caption{{\label{Fig_A1}}Convergence of the first 20 (including
  Kramers degeneracy) electron energy levels ${\cal E}^e_n$ of dots with
  different heights as a function of {\bf k}-point meshes and for different
  supercell sizes [$(L_x{\times}L_y{\times}L_z)$]. The results are color coded
  to help clarity, and are given w.r.t. GaAs CBM [$E_c(GaAs)=-4.093\,{\rm
    eV}$]. The basis encompass single-band, $\Gamma_{1c}$ GaAs and
  strained-InAs bulk Bloch functions in the vicinity of the $\Gamma$ point of
  the supercell. The 3-fold $(Q,Q,K)$ represents the main-axis sizes of an
  ellipsoid in reciprocal space. This ellipsoid determines the number of ${\bf
    k}$ points in the basis, which is the {\em same} for
  GaAs and InAs.}
\end{figure*}

{\em Supercell size.---}We consider cubic supercells of size
$L_x{\times}L_y{\times}L_z$, where $L_{t}$ ($t=x,y,z$) is measured in units of
the lattice parameter of bulk GaAs $a_{GaAs}$. Based on strain calculations
(see above), we selected a supercell size such that the strain profile within
the quantum dot remained unchanged upon changing size.  Notwithstanding, it
should be noted that although the strain may be converged within the dot the
strain values at the boundary of the supercell may not be zero due to the slow
relaxation of the strain fields. This residual strain introduces small changes
($\sim{\rm meV}$) in the energy levels. For instance, Figure \ref{Fig_1} shows
the effect of reducing the residual strain by changing the supercell size from
$72{\times}72{\times}40$ to $72{\times}72{\times}48$ for a $h=20\,${\AA} dot.

{\em Number of {\bf k} points.---}The {\bf k}-point meshes enclose all the
${\bf k}$ vectors around the $\Gamma$ point that lie within an ellipsoid with
main axis equal to $(2\pi/L_x)\,P$, $(2\pi/L_y)\,Q$, and $(2\pi/L_z)\,K$,
respectively. P, Q, and K give the number of {\bf k} points taken along each
cartesian direction. In our convergence assesment $P=6,\;8$, $Q=P$, and $K$
ranges from 8-18. Figure \ref{Fig_A1} shows the electron energy levels for
several dots as a function of {\bf k}-point mesh. We distinguish several
features. (a) When compared with the energy of the levels, differences in
energies are quicker to converge. (b) High-energy levels require larger {\bf
  k}-point meshes to converge within a given threshold. (c) The taller the
quantum dots the smaller the number of {\bf k} points needed to converge. (d)
The use of $(6,\,6,\,K)$ and $(8,\,8,\,K)$ {\bf k}-point meshes results in
similar energies for the lowest electron level. The discrepancy between the
predictions made with these two meshes increases for high-energy levels.

\begin{figure}
\includegraphics[width=7.5cm]{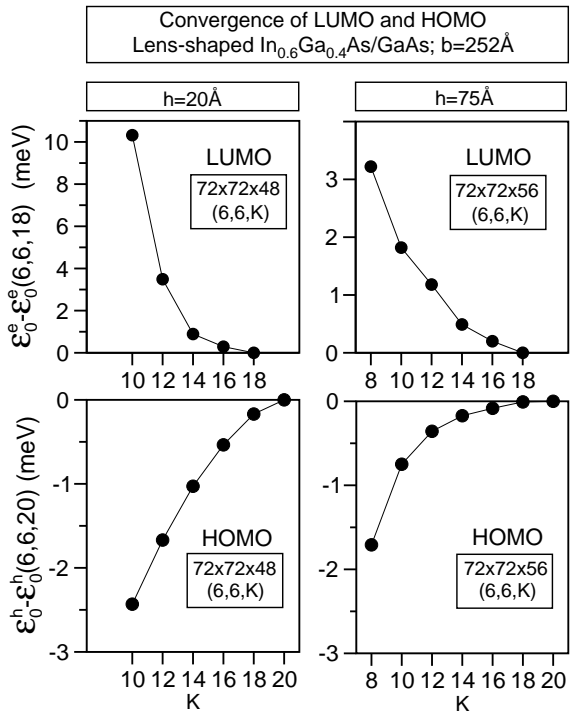}
\caption{{\label{Fig_A2}}Convergence of LUMO (${\cal E}^e_0$) and HOMO (${\cal E}^h_0$) energies relative to their 
  value simulated with the highest number of {\bf k} points on a basis of the
  form $(6,\,6,\,K)$. To simulate the HOMO energy we include the 3
  $\Gamma_{15v}$ bands. The size of the supercell is indicated.}
\end{figure}

To present details of the convergence with respect to {\bf k}-point mesh,
Figure \ref{Fig_A2} shows a convergence assessment of HOMO and LUMO's energy
in flat ($h=20\,${\AA}) and tall ($h=75\,${\AA}) dots. The energy of HOMO
converges visibly quicker than that of LUMO.  This behavior holds for all hole
energy levels, and arises from the higher size (three $\Gamma_{15v}$ bands) of
the basis used in the simulation.

%

To summarize, the results we present in this work for each quantum dot derive
from simulations with {\bf k}-point meshes that provide with energy levels
converged within $2{\rm meV}$. Table \ref{Table_1} shows the supercell sizes
and {\bf k}-point meshes we have used to simulate the electronic structure of
each of the dots we have studied.  It should be noted that {\em energy
  differences} are converged to much less than this lower bound.

\begin{table}
\caption{{\label{Table_1}}Summary of supercell sizes
  $L_x{\times}L_y{\times}L_z$ (lengths in units of $a_{GaAs}$) and {\bf k}-point meshes $(Q,\,Q,\,K)$ used in
  the simulations. Dot's base $b=252\,${\AA}. (c.f. Fig. \ref{Fig_1} for dot's geometry.)}
\begin{tabular}{ccc}
\hline\hline
height ({\AA})& $L_x{\times}L_y{\times}L_z$ & {\bf k}-point mesh \\ 
\hline 
 20 & $72{\times}72{\times}48$ & $(6,\,6,\,13)$ \\
 35 & $72{\times}72{\times}48$ & $(6,\,6,\,13)$ \\
 50 & $72{\times}72{\times}48$ & $(6,\,6,\,12)$ \\
 65 & $72{\times}72{\times}48$ & $(6,\,6,\,12)$ \\
 75 & $72{\times}72{\times}56$ & $(6,\,6,\,10)$ \\
\hline\hline
\end{tabular}
\end{table}

%

%
\end{document}